\documentclass[a4paper,fleqn]{cas-dc}

\usepackage[numbers]{natbib}
\graphicspath{{figs/}{./}}

\def\tsc#1{\csdef{#1}{\textsc{\lowercase{#1}}\xspace}}
\tsc{WGM}
\tsc{QE}
\tsc{EP}
\tsc{PMS}
\tsc{BEC}
\tsc{DE}

\begin{document}
\let\WriteBookmarks\relax
\def\floatpagepagefraction{1}
\def\textpagefraction{.001}
\shorttitle{Signal or Noise?}
\shortauthors{A. Georgakopoulos et~al.}

\title [mode = title]{Signal or Noise? Modality Contribution and Cooperation in Multimodal GraphRAG}

\author[1]{{Antonios Georgakopoulos}}
\cormark[1]
\ead{a.georgakopoulos@uva.nl}
\credit{Writing - Original Draft, Conceptualization, Methodology, Software, Validation, Formal analysis, Investigation, Data Curation, Visualization}

\affiliation[1]{organization={University of Amsterdam},
                city={Amsterdam},
                country={The Netherlands}}

\author[1]{{Paul Groth}}[orcid=0000-0003-0183-6910]
\ead{p.t.groth@uva.nl}
\credit{Writing - Review \& Editing, Conceptualization, Methodology, Supervision}

\author[1]{{Lise Stork}}[orcid=0000-0002-2146-4803]
\ead{l.stork@uva.nl}
\credit{Writing - Review \& Editing, Conceptualization, Methodology, Supervision, Funding acquisition}

\cortext[cor1]{Corresponding author}

\begin{abstract}
Multimodal knowledge graphs (KGs) integrate information from text, figures, tables, and other modalities into a unified structured representation, with the promise that richer evidence enables better inference. In GraphRAG systems built over such graphs, it is commonly assumed that retrieving evidence from more modalities at inference time improves downstream performance. Yet, redundant or overlapping multimodal evidence may distract language models in question answering (QA), and whether each modality contributes equally across questions, models, and tasks remains poorly understood. In this work, we study how modality-aware retrieval affects downstream inference in a multimodal GraphRAG pipeline, using document visual question answering (DocVQA) as a testbed. We extend an existing KG-based QA framework to be modality-aware, leveraging the graph structure to track which modality supports which facts and to selectively filter evidence at the edge level. This enables us to investigate whether providing all available multimodal evidence at inference time benefits QA, and to evaluate the contribution and cooperation of modalities across question, task, and model characteristics. Through a controlled analysis within a state-of-the-art multimodal GraphRAG pipeline, five multimodal LLMs and two DocVQA benchmarks, we find that tables and text provide the strongest contributions, and that combining modalities frequently produces redundancy rather than synergy, particularly for pairs involving textual information. Positive cooperation appears mainly between non-text modalities and depends on question intent and task type. Our findings argue for selective, modality-aware retrieval in the design of more effective GraphRAG systems, where modalities are filtered according to the downstream task rather than retrieved uniformly.

\end{abstract}




\maketitle

\section{Introduction}

\begin{figure*}
  \centering
  \includegraphics[width=\textwidth]{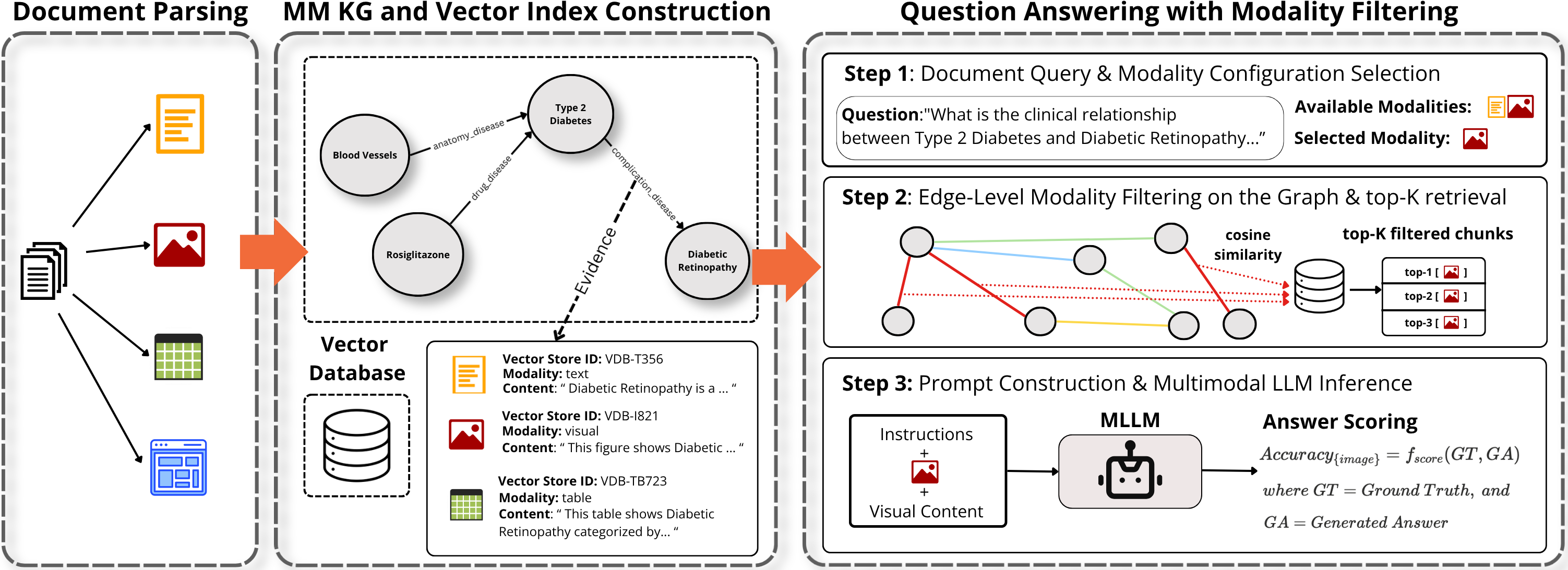}
  \caption{Overview of our modality-aware multimodal QA framework.}
  \label{fig:pipeline}
\end{figure*}

GraphRAG systems increasingly leverage multimodal knowledge graphs (KGs) that integrate information from text and other modalities into a unified representation for retrieval and inference~\cite{bu2025query}. A common assumption in these systems is that incorporating additional modalities improves downstream performance~\cite{riedler2024beyond,peng2025unidoc,mkg_rag_yuan,suri2025visdom}. However, the individual contributions of different modalities on the downstream task, as well as whether they provide complementary or redundant evidence during inference, remain underexplored.
Prior work in multimodal learning has investigated modality dominance, robustness, and modality collapse~\cite{gapp2025you,mai2023multimodal,liang2022mind}. In multimodal RAG, recent benchmarks such as \textit{mmRAG}~\cite{xu2025mmrag} enable modality-specific evaluation, and several studies report that combining heterogeneous evidence improves QA performance~\cite{riedler2024beyond,peng2025unidoc,mkg_rag_yuan,suri2025visdom}. Nevertheless, most existing analyses rely on modality ablation experiments, which provide limited insight into how modalities interact during inference across different models, domains, question intents, and reasoning tasks.
Understanding modality interactions is important because each modality may contribute distinct forms of evidence during question answering. For example, medical images may capture visual indicators of disease~\cite{wu2025mkgf,xia2024rule}, while tables may better encode structured quantitative information in financial QA~\cite{zhao2025finragbench}. Similarly, the usefulness of a modality may depend on the question type: verification questions may rely primarily on textual evidence, whereas calculation or counting questions may benefit more from tabular or visual information.
Multimodal GraphRAG systems provide a useful setting for studying these interactions because they represent evidence from different modalities as interconnected graph entities and relations. This graph structure makes it possible to trace which modalities support individual facts and to analyse how modality specific evidence contributes to downstream retrieval and question answering. Such an analysis has direct implications for GraphRAG: it can show when evidence from multiple modalities is beneficial, and when additional modalities provide little or even negative value. 

Therefore, our contributions are as follows:
\begin{enumerate}
    \item We propose an extention to a state-of-the-art multimodal GraphRAG system that allows for the tracing of each KG triple back to the specific modality (and part of the document) that it was extracted from. 
    \item We propose a modality contribution evaluation framework that tests modality configurations, using Shapley Value-based metrics that describe: (1) how much each modality contributes to the overall performance, and (2) if the modalities provide a synergistic or redundant effect.
    \item We perform an analysis using two DocVQA benchmarks and five multimodal LLMs to understand how different modalities and modality combinations impact question-answering performance across question intents, task types, and inference models.
\end{enumerate}

\section{Related Work}

RAG systems augment LLMs~\cite{lewis2020retrieval} with information retrieved from external knowledge sources, while multimodal RAG extends this paradigm to multimodal data~\cite{chen-etal-2022-murag}.

\subsection{Modality contributions in multimodal RAG.}
Prior work has shown that combining modalities often improves retrieval and QA performance. Riedler et al.~\cite{riedler2024beyond}, Peng et al.~\cite{peng2025unidoc}, and VisDoM~\cite{suri2025visdom} report that multimodal retrieval generally outperforms unimodal settings. CMRAG~\cite{chen2026cmragcomodalitybasedvisualdocument} shows that the advantages of multimodal retrieval depend on how the modalities are combined. The authors find that naively combining textual and visual information performs worse then using a single modality. VisRAG~\cite{visrag} introduces a RAG pipeline that utilises a VLM to read document pages directly and perform retrieval and answer generation. This VLM-based RAG approach outperforms text-based RAG pipelines on answer generation. Recent multimodal GraphRAG approaches further suggest synergistic effects between modalities. For example, Bu et al.~\cite{bu2025query} show that combining visual evidence and tabular relationship mining yields larger gains than using either individually. Similarly, Yuan et al.~\cite{mkg_rag_yuan} introduce a multimodal GraphRAG approach that aligns text and image content in the same graph and outperforms text-only and vision-only variants of the same pipeline.

\subsection{Modality robustness in multimodal learning.}
Several works study modality reliability and dominance in multimodal models. Mai et al.~\cite{mai2023multimodal} show that reweighting modalities based on reliability improves robustness, while Liang et al.~\cite{liang2022mind} demonstrate that modality separation in embedding spaces affects downstream performance. Gapp et al.~\cite{gapp2025you} further show that some multimodal architectures rely predominantly on a single modality. Another study, by Wang et al.~\cite{wang2026compose}, shows that in logical reasoning tasks using a second modality helps only when the modality can provide useful information about the answer, otherwise performance often degrades. Peng et al.~\cite{peng2022balanced} demonstrate that the dominance that a modality can have in a multimodal setting may arise during training. Specifically, the modality that learns faster can absorb the shared gradient and leave the other modality under-trained. Wei et al.~\cite{wei2024enhancing} show that dominance of a modality varies across samples in the dataset, and so methods that correct imbalance on the dataset-level can fail when the dominant modality differs from sample to sample.

\subsection{Metrics for modality contribution analysis.}
Prior work has proposed methods for quantifying modality importance and interactions. MMSHAP~\cite{parcalabescu2023mm} uses Shapley values to measure modality contribution and multimodal collapse. Amit et al.~\cite{amit2025quantifying} argue that standard ablations conflate importance and interaction effects, and propose a partial information decomposition (PID) framework to separate unique, redundant, and synergistic contributions. Yang et al.~\cite{yang-quantification} also introduce a PID method that studies the interactions on the sample level, showing that how modalities interact varies across samples. Gat et al.~\cite{gat2021perceptual} introduce the perceptual score that measures how much a model relies on each modality by permuting that modality during test time and observing how much the performance drops. The authors in~\cite{wei2024enhancing} use a Shapley value-based metric to evaluate the contribution of each modality for each sample of the dataset. SHAPE~\cite{ijcai2022p425} metric uses Shapley values to quantify both marginal modality contribution and cross-modal cooperation. We adopt SHAPE in this work because it captures both individual modality importance and modality interactions.

Despite growing interest in multimodal RAG, most prior work evaluates modality effects through retrieval comparisons or ablations, in which a modality is removed and the resulting change in performance is reported. These ablations measure the gain in performance from a modality in the presence of the other modalities, giving us a single number. This number combines the usefulness of the modality with its interaction with the remaining modalities. SHAPE however evaluates each modality under all combinations of the available modalities and provides two separate quantities from them. The first is a contribution score that averages the gain provided by the modality with and without the other modalities present. The second is a cooperation score which captures if the joint use of the modalities yields more than the sum of their individual contributions. Therefore, ablation-based studies provide limited insights into whether modalities contribute complementary or redundant evidence. A large performance drop when a modality is removed may indicate either that the modality carried unique information or that it was only useful in combination with another modality. Our work addresses this gap by analyzing modality contributions and cooperation in multimodal GraphRAG systems using SHAPE~\cite{ijcai2022p425}.

\section{Task Definition}
\label{problem_definition}
Let $\mathcal{D}$ be a collection of documents. Each document $d \in \mathcal{D}$ contains content from multiple modalities. We denote the set of modalities by $\mathcal{M} = \{\text{text}, \text{table}, \text{image}, \text{layout}\}$.
Each document is represented as a multimodal knowledge graph 
\[
G_d = (V_d, E_d)
\]
where $V_d$ is the set of nodes, corresponding to elements in the document content, and associated with a modality label $m \in \mathcal{M}$ corresponding to the type of content, and $E_d$ refers to the set of edges. 

We define the task of \textit{modality-aware question answering}, where inference is conditioned on a subset of modalities in multimodal knowledge graph $G_d$. Given a question $q$ about a document $d$, the goal is to produce an answer $a$ by retrieval and inference over $d$ scoped by a subset of the graph $G_d$. In a modality-aware setting, retrieval for question answering is constrained to a subset of modalities $M \subseteq \mathcal{M}$. To enable modality-specific retrieval, we associate each fact in the graph, i.e. each edge $(e_s, r, e_t) \in E_d$, with the set of modalities of the document content from which it was extracted.

\section{Modality-Aware Framework}\label{sec:framework}

We build on an existing state-of-the-art KG-based question answering framework and extend it to support modality-aware retrieval and analysis, see Figure~\ref{fig:pipeline}. Our approach introduces modality-aware provenance tracking and uses it to constrain retrieval during question answering. 

\textit{Modality Tracking:}
We add a mechanism to track which modality supports each fact represented in the graph. More specifically, for each edge $(e_s, r, e_t)$ on the graph we define an evidence set $\mathcal{E}_d(e_s, r, e_t) \subseteq X_d$ as the chunks from which the edge was extracted, where $X_d = \{x_1, \dots, x_N\}$ denotes the set of all chunks in document $d$. Since each chunk $x_i$ is associated with a modality label $m(x_i)$, we define the modality set for this edge as:
\[
M_d(e_s,r,e_t) = \{\, m(x_i) \mid x_i \in \mathcal{E}_d(e_s,r,e_t) \,\}
\]
This gives each edge a modality provenance record, allowing us to filter graph edges based on the modalities of their supporting evidence.

\textit{Modality-Aware Retrieval:}
\label{qa_section} 
To analyse modality-aware question answering, we answer each question $q$ by querying $G_d$ under different modality subsets $M \subseteq \mathcal{M}$. For a given modality subset $M \subseteq \mathcal{M}$, we restrict the graph to edges $(e_s, r, e_t) \in E_d$ whose supporting modalities satisfy $M_d(e_s, r, e_t) \cap M \neq \emptyset$. For each retained edge we collect the subset of its evidence set $\mathcal{E}_d(e_s, r, e_t)$ drawn from those modalities, so a fact supported by several modalities is kept, but only its chunks from $M$ are used. The context $\text{c}_{q,M}$ is formed from the retained edges, specifically the entities and relations whose supporting evidence falls within the supporting evidence, together with their supporting chunks. A large language model (LLM) is then prompted with this context to answer question $q$.

\section{Experimental Setup}
We now discuss our experimental setup beginning with the GraphRAG QA system and the associated LLMs used. We then describe the benchmarks employed and how we categorise questions. Lastly, evaluation metrics and the analyses we perform are then detailed. 

\subsection{GraphRAG QA System}
In this work, we employ the RAG-Anything pipeline~\cite{guo2025rag} as our graph-based multimodal RAG framework, as it achieves state-of-the-art performance on multimodal document question answering benchmarks. However, our framework is model-agnostic and can be applied to other GraphRAG QA systems. RAG-Anything constructs the graph in two stages: (1) decomposing the document into modality-specific content units, each carrying a modality label, and (2) extracting entities and relations from these units to build a unified knowledge graph. We modify the graph construction stage of the pipeline so that modality-aware provenance is recorded on the edges of the graph. This enables our modality-specific retrieval, as follows. Given a modality subset $M$, we restrict $G_d$ to the edges whose supporting modalities intersect $M$ and collect their supporting chunks of those modalities. These chunks are then ranked by the embedding similarity of their content to the question $q$, and the top-k form the context $\text{c}_{q,M}$. An MLLM is then used to construct an answer $a$, given the context.

\subsection{LLMs for Answer Generation}

For answer generation, we evaluate five multimodal LLMs with different model families and numbers of parameters. Specifically, we employ GPT-4o-mini~\cite{menick2024gpt}, two Qwen3-VL~\cite{bai2025qwen3} models with 8B and 30B parameters, and two Gemma 3~\cite{team2025gemma} models with 4B and 27B parameters. This selection allows us to compare modality contribution and cooperation patterns across both proprietary and open-weight models, as well as across smaller and larger model variants within the same family. These five models are used as the answer-generation models in our QA pipeline. For the retrieval component, we use text-embedding-3-large to compute embeddings for both document chunks and questions, which are then used for similarity search over the vector database.

\subsection{DocVQA Benchmarks}
\label{datasetandquestionselection}

\begin{table}[b]
\centering
\scriptsize
\setlength{\tabcolsep}{4pt}
\caption{Number of retained two-modality questions per benchmark and modality pair.}
\label{tab:two_modality_question_counts}
\begin{tabular}{lrrr}
\toprule
Modality Pair & LongDocURL & MMLongBench-Doc & Total \\
\midrule
Image + Layout & 34  & 47 & 81 \\
Image + Table  & 164 & 16 & 180 \\
Image + Text   & 152 & 92 & 244 \\
Layout + Table  & 150 & 9  & 159 \\
Layout + Text   & 280 & 16 & 296 \\
Table + Text    & 55  & 36 & 91 \\
\midrule
Total & 835 & 216 & 1,051 \\
\bottomrule
\end{tabular}
\end{table}

Experiments are conducted using two DocVQA benchmarks: (1) the MMLongBench-Doc \cite{ma2024mmlongbench} and (2) LongDocURL~\cite{deng2025longdocurl}. Both benchmarks are long-context multimodal document understanding benchmarks that contain multi-page documents from diverse domains, with information distributed across text, figures, and tables. 
We select these benchmarks as they provide gold-standard annotations of the modalities required to answer each question. This is essential for our analysis, as we evaluate how the answer of a question changes when restricting the retrieved evidence to subsets of the modalities required for the question. The two benchmarks together contain 327 documents spanning multiple domains, such as research reports, financial reports, brochures, etc. Documents vary substantially in length and structure, and many questions require cross-page reasoning.

The evidence sources in MMLongBench-Doc include \textit{Pure-text, Generalized-text (Layout), Chart, Figure}, and \textit{Table}, whereas for LongDocURL the evidence sources include \textit{Text, Layout, Figure}, and \textit{Table}. Since both benchmarks use closely related modality definitions, we map them into a shared modality space consisting of \textit{Text, Layout, Image}, and \textit{Table}. For MMLongBench-Doc, we thus merge \textit{Chart} and \textit{Figure} into the \textit{Image} modality, since both correspond to visual evidence. For LongDocURL, we treat \textit{Figure} as \textit{Image}, allowing us to maintain a consistent modality convention across benchmarks. For each question, the gold modality set $M_{\text{gold}}(q)$ is constructed using this mapping. 

For our experiments, we focus on questions whose gold modality set contains exactly two modalities, resulting in 1,051 questions. This decision enables controlled evaluation by comparing performance across all non-empty modality subsets $M \subseteq M_{\text{gold}}(q)$. Questions whose evidence is found only on a single modality are excluded, as they do not permit measuring how modalities interact. In both benchmarks only a small number of questions, 41 in total, have three supporting modalities, but these are too sparse to analyse reliably as the largest three-modality subgroup contains only 26 questions. We therefore restrict the analysis to the two-modality case. Each question is evaluated under all three non-empty subsets of its gold modality set and across all five MLLMs, resulting in a total of \(1,051 \times 3 \times 5 = 15,765\) question-answering runs. Table~\ref{tab:two_modality_question_counts} shows the number of retained questions and their corresponding modalities for each benchmark.

\subsection{Question Intent Taxonomy}\label{sec:taxonomy}

To classify benchmark questions according to intent (wording + what a question intends to ask), we adopt a web search question intent taxonomy~\cite{cambazoglu2021intent} as a proxy for our setting. This taxonomy was designed for natural-language questions, and constructed based on real user questions. Although DocVQA and web search are different tasks, both tasks involve questions that seek information and must be grounded in external evidence. 
Rose and Levinson~\cite{rose2004understanding}, for example, argue that searching is a means of satisfying an underlying user goal and organize search goals around what the user is trying to accomplish. Table~\ref{tab:intent_categories} lists the intent categories that we used for our analysis as offered by the taxonomy.

\begin{table*}
\caption{Intent categories from the question intent taxonomy~\cite{cambazoglu2021intent}.}
\centering
\scriptsize
\setlength{\tabcolsep}{4pt}
\renewcommand{\arraystretch}{1.15}
\begin{tabular}{p{0.11\textwidth} p{0.36\textwidth} p{0.11\textwidth} p{0.36\textwidth}}
\hline
\textbf{Type} & \textbf{Description} & \textbf{Type} & \textbf{Description} \\
\hline

Description & This category covers mostly ``what is X'' or ``what is X of Y'' questions, where the user aims to obtain a definition or description of an object or one of its attributes.
&
Quantity & These questions expect a numeric value as answer (e.g., price, frequency, duration, speed, age, length, weight)
\\
\hline

Process & These are typically ``how to do X'' questions that seek instructions, guidelines, or procedures which will facilitate an action to be performed by the user in real life.
&
Entity & The expected answer to a question in this category is a named entity, excluding numeric entities.
\\
\hline

Advice & This category includes questions where the user aims to obtain personal advice on a particular topic. The Advice category differs from Process in that the former expects a somewhat subjective recommendation, whereas the latter expects an objective, step-by-step process description.
&
Language & These questions usually provide a name or description, and the expected answer is another name or object (e.g., ``how is X called'', ``what is X in the Y language''). Questions seeking a named entity as answer are excluded.
\\
\hline

Opinion & These are questions seeking to get a subjective opinion about a topic of interest (e.g., ``what do you think about X'' or ``is X good/bad''). This class excludes Advice questions, where the subject is the user issuing the question.
&
Temporal & These are typically the ``when is X'' questions, which expect to obtain a date or time of an event as answer.
\\
\hline

Verification & These are fact-checking questions that seek an affirmative yes/no answer which cannot be disputed.
&
List & The expected answer is an itemized list whose items can have any type (e.g., entity, quantity, or a mix). For example, ``which countries won the world cup'' is a List question.
\\
\hline

Attribute & ``what is Y of X'' questions that seek a particular property of a given named entity are in this category. We exclude questions whose answers are named entities to avoid a possible overlap with the Quantity and Entity categories.
&
Calculation & These are questions aiming to use the search engine as a calculator for arithmetic operations or unit conversion. Similar to the Quantity category, the expected answer is a numeric value, but the Calculation category contains one or more numeric values as input in the question.
\\
\hline

Reason & The expected answer to these questions include an explanation of causes underlying a certain action or event. Most questions of type ``why is/do X Y'' are in this category.
&
Weather & These are questions about weather forecast.
\\
\hline

Location & These are typically ``where is X'' questions that seek the position, address, or location of a given object or entity. The answers are not limited to geo-locations. For example, ``where is X in human body'' falls in this category.
&
Resource & Non-informational questions where the goal is to obtain an online or offline resource are in this category, excluding questions with Calculation and Weather intent
\\
\hline

\end{tabular}

\label{tab:intent_categories}
\end{table*}

\begin{figure}
  \centering
  \includegraphics[width=\columnwidth]{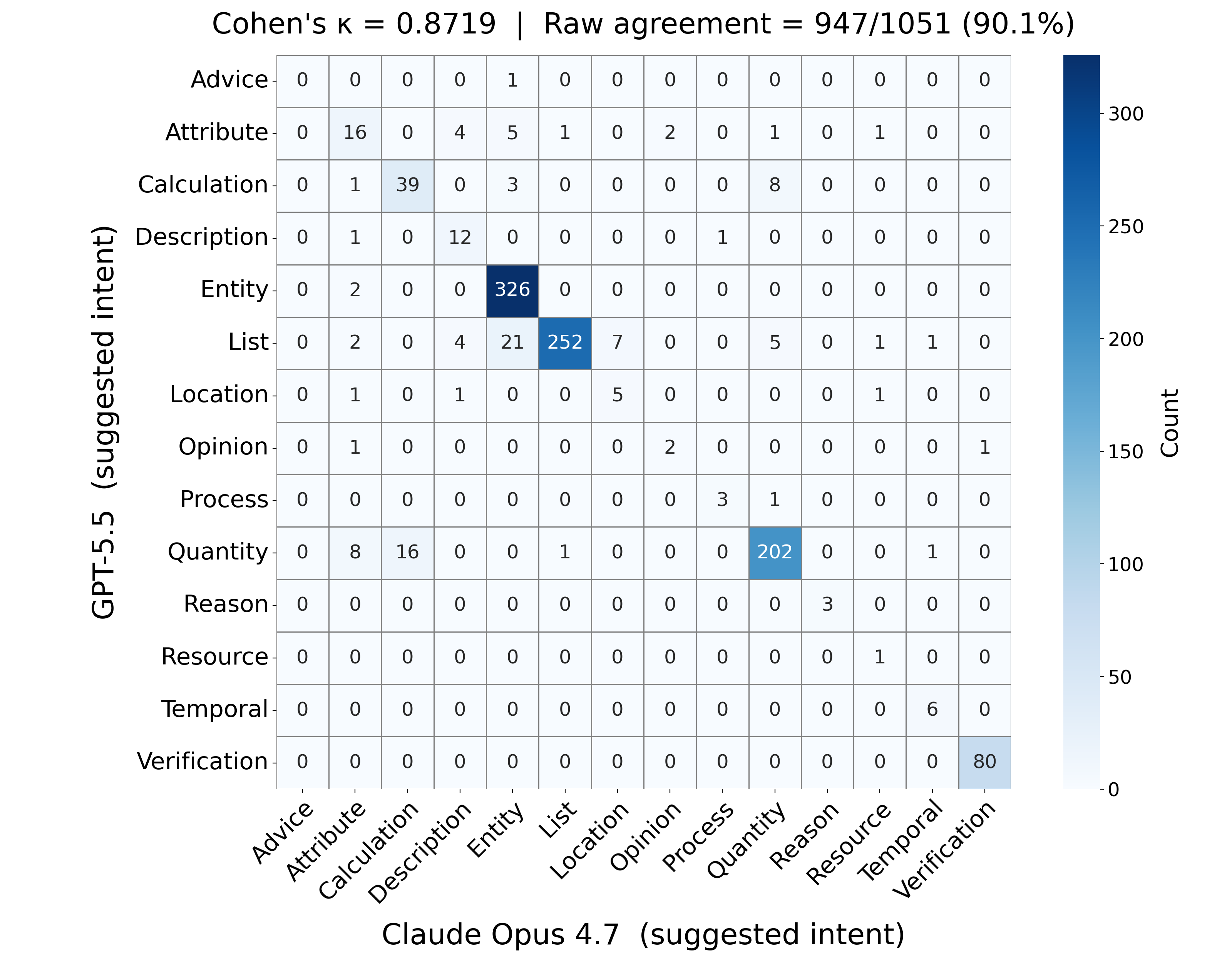}
  \caption{Confusion matrix comparing the final intent labels assigned by GPT-5.5 and Claude Opus 4.7 during the judging stage. Rows correspond to GPT-5.5 labels and columns correspond to Claude Opus 4.7 labels.}
  \label{fig:confusion_matrix_intent}
\end{figure}

We map every benchmark question to each category from the Intent Taxonomy. To do this we follow a zero-shot intent classification prompting approach~\cite{parikh2023exploring}, which has been shown to perform competitively on this task, and use the GPT-5 model~\cite{openai_gpt5} for classifying all questions. To judge the quality of these annotations, we use two frontier LLM models, GPT-5.5~\cite{openai_gpt55} and Claude Opus 4.7~\cite{anthropic_opus47}, to act as judges and validate the annotations. They either agree with the given label, or propose an alternative one from the taxonomy. We change the label of a question only when both judges agree on the proposed label. The prompts of both the intent classification and the judging step can be found in our GitHub repository\footnote{\url{https://github.com/Antonis-Georgakopoulos/mmkg-modality-analysis/blob/main/question_intent_analysis/question_intent_classification/prompt.txt}}\footnote{\url{https://github.com/Antonis-Georgakopoulos/mmkg-modality-analysis/blob/main/question_intent_analysis/question_intent_classification/judge_prompt.txt}}. To quantify agreement between the two LLM judges, we compute Cohen's kappa coefficient which measures the inter-annotator agreement for categorical labels~\cite{cohen1960coefficient}. Alignment between the judges was high, with Cohen's $kappa = 0.87$, and a raw agreement of $90.1\%$. GPT-5.5 flagged 97 out of 1,051 questions, and Opus 4.7 146 questions, with both models identifying the same 73 questions. Of those questions, the models agreed on the replacement intent of 66 questions, and these consensus revisions were applied. We report main agreement statistics in Table~\ref{tab:llm_judge_agreement} and provide the full confusion matrix between the two LLM judges in Figure~\ref{fig:confusion_matrix_intent}.  

\begin{table}[t]
\centering
\scriptsize
\caption{Agreement between two LLM judges during question intent-label validation.}
\label{tab:llm_judge_agreement}
\begin{tabular}{lr}
\toprule
\textbf{Measure / outcome} & \textbf{Value} \\
\midrule
Total questions & 1,051 \\
Final-label agreement & 947 / 1,051 (90.1\%) \\
Cohen's $\kappa$ & 0.8719 \\
\midrule
\multicolumn{2}{l}{\textit{Decision on initial annotation}} \\
Both judges accepted the initial annotation & 881 \\
Only GPT-5.5 suggested a change & 24 \\
Only Claude Opus 4.7 suggested a change & 73 \\
Both judges suggested a change & 73 \\
\midrule
\multicolumn{2}{l}{\textit{Agreement when both judges suggested a change}} \\
Same replacement intent & 66 / 73 (90.4\%) \\
Different replacement intent & 7 / 73 (9.6\%) \\
\bottomrule
\end{tabular}
\end{table}

\subsection{Evaluation Metrics}
\label{sec:metrics}

\subsubsection{Answer Accuracy}
We first evaluate generated answers using the accuracy metric as defined in the benchmarks. This produces a per-question score $v$.

\subsubsection{SHAPE Metric}
To quantify modality contributions, we adopt the SHAPE metric~\cite{ijcai2022p425}, a Shapley value-based evaluation metric. Let $V(M)$ denote the aggregated performance over a set of questions when the system is restricted to using modalities in $M \subseteq \mathcal{M}$. SHAPE computes the marginal contribution of each modality to the performance function $V$, and the degree of cross-modal cooperation between modalities. 

\paragraph{Modality contribution $S$.}

Following the SHAPE metric, the contribution $S$ of modality $m_1$ is defined as:
\[
S_{m_1} = \frac{1}{2} \cdot \frac{
V(\{m_1, m_2\}) - V(\{m_2\}) + V(\{m_1\}) - V(\emptyset)
}{Z_f}
\]
where $Z_f = V(m_1,m_2)$ is the normalization factor. 

\paragraph{Cross-modal cooperation $C$.}

The cooperation between modalities $m_1$ and $m_2$ is defined as: 
\[
C_{\{m_1,m_2\}} =
V(\{m_1, m_2\}) - V(\{m_1\}) - V(\{m_2\}) + V(\emptyset).
\]

A positive cooperation score indicates complementary behaviour, where the joint use of the modalities yields higher performance than the sum of their individual contributions. A score close to zero suggests additive behaviour with limited interaction. A negative cooperation score indicates redundancy or interference, where the combined performance is lower than the sum of the individual contributions.

\paragraph{Baseline value $V(\emptyset)$.} Both formulas require a baseline value $V(\emptyset)$, which is the performance that we get when no modality evidence is available. SHAPE metric uses a majority-class predictor as this baseline compared to null inputs, so we adopt the same approach for generative QA. For each modality group we take the most frequent gold answer among the group's questions, predict it for every question in the group, and score according to the evaluation of each benchmark.

\paragraph{Contribution score difference $D$.} We report the pairwise contribution difference as:
\[
D_{S_{m_2},S_{m_1}} = S_{m_2} - S_{m_1},
\]
where $S_{m_1}$ and $S_{m_2}$ are the SHAPE contribution scores of the two modalities in a pair. This value allows us to ask whether, for a given pair, one modality consistently contributes more to the task performance than the other. Positive values of $D$ indicate that $m_2$ contributes more than $m_1$, while negative values indicate the opposite. 

\subsubsection{Bootstrap Sampling Schemes}

We use bootstrapping to quantify uncertainty and assess variability across different data samples, and employ the following main and additional bootstrapping schemes: 
\begin{enumerate}
\item \textbf{Stratified bootstrapping resampling (main):}, stratified resampling which preserves modality group distribution within each benchmark; 
\item \textbf{Modality pair-only stratified bootstrapping:} stratified resampling within each modality group, pooling questions across benchmarks;  
\item \textbf{Unstratified bootstrapping:} resampling from the fully pooled set of questions, ignoring both benchmark and modality-group structure.
\end{enumerate}
 
In all analyses, we use bootstrap resampling with replacement and report percentile-based 95\% confidence intervals. We draw 100,000 bootstrap samples, shared across all models to ensure fair comparisons. We additionally report the mean pairwise Spearman correlation $\bar{\rho}s$ across bootstrap samples and the cross-model mean $\mu$. All bootstrap results in Section~\ref{sec:model-analysis} report the main sampling scheme, since additional bootstrap schemes produced largely consistent confidence intervals, indicating robustness to the sampling strategy.

\subsection{Experimental Analyses}

To understand modality-aware behavior, we perform the following analyses: 

\begin{enumerate}
    \item \textbf{Overall Performance} We report performance scores across all questions, benchmarks and models. These performance scores provide a reference for the SHAPE analysis, as the SHAPE scores should be considered in context of overall model performance. 
     \item \textbf{Model Comparison} We compare contribution $S$ and cooperation $C$ scores across different LLMs to assess how model choice affects the utilization of information from different modalities. For robustness, all aggregate comparisons use bootstrap confidence intervals as described in Section~\ref{sec:metrics}; 
    \item \textbf{Question-Type Analysis} We compare contribution $S$ and cooperation $C$ scores across different question types to understand how modality contributions vary with question \textit{intent} (see Section~\ref{sec:taxonomy} for a description of our question classification method and taxonomy) and the processes it requires to answer it (\textit{process type}). 
\end{enumerate}

\section{Results}

\begin{figure*}
  \centering
  \includegraphics[width=\linewidth]{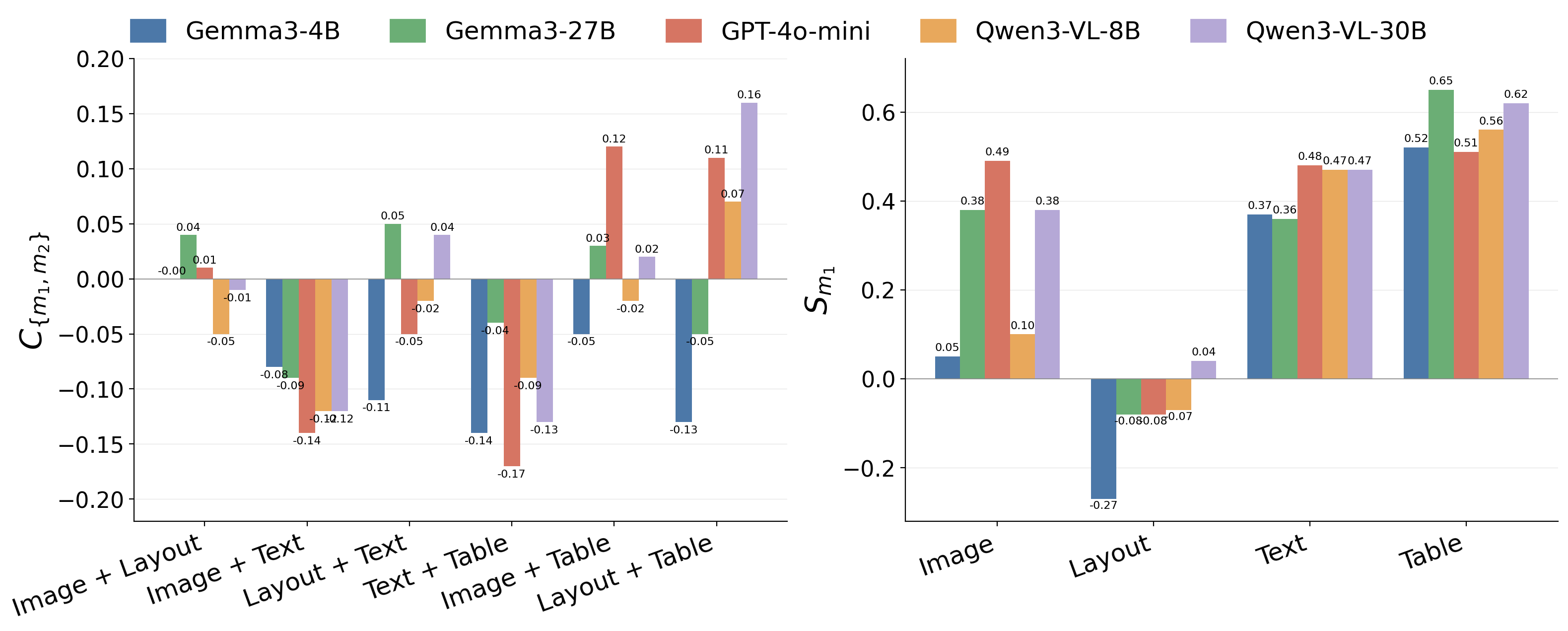}
  \caption{SHAPE cooperation scores $C_{\{m_1,m_2\}}$ for each modality pair, and contribution scores ($S_{m_1}$) macro-averaged across the two benchmarks for every model. For cooperation scores, values indicate complementary benefits when the two modalities are used together, values close to zero indicate limited interaction, and negative values indicate that their joint performance is lower than the sum of their individual contributions. For contribution scores, higher values indicate a larger marginal contribution to QA task performance.}
  \label{fig:aggregated_results_per_model_C_and_S}
\end{figure*}

We first produce per-question scores $v$, as a basis for our SHAPE (contribution and cooperation) analyses. 
The per-benchmark and per-model accuracy scores can be found in Table~\ref{app:tab:overall_accuracy}. Results regarding the absolute accuracy differ between the benchmarks, however analysing the observed differences in accuracy is beyond the scope of this work.

\begin{table}[t]
\centering
\small
\setlength{\tabcolsep}{2.5pt}
\caption{Total accuracy (\%) and accuracy by modality group across benchmarks.The value $n$ denotes the number of evaluated questions in each modality group. We use the following model abbreviations: G3-27B = Gemma3-27B, G3-4B = Gemma3-4B, GPT = GPT-4o-mini, Q3-30B = Qwen3-VL-30B, and Q3-8B = Qwen3-VL-8B.}
\label{app:tab:overall_accuracy}

\begin{tabular}{@{}lrrrrrr@{}}
\toprule

\multicolumn{7}{c}{\textbf{LongDocURL}} \\
\textbf{Modality Group} & $n$ & G3-27B & G3-4B & GPT & Q3-30B & Q3-8B \\
\midrule
Image + Layout & 34  & 23.34 & 26.74 & 45.36 & 54.90 & 50.21 \\
Image + Text   & 152 & 41.13 & 28.10 & 52.77 & 59.14 & 58.10 \\
Image + Table  & 164 & 13.97 & 6.86  & 33.41 & 37.95 & 39.05 \\
Layout + Text  & 280 & 34.86 & 22.25 & 47.28 & 43.64 & 49.81 \\
Layout + Table & 150 & 17.69 & 12.71 & 32.29 & 35.57 & 37.10 \\
Text + Table   & 55  & 50.10 & 28.23 & 64.51 & 78.40 & 74.39 \\
\cmidrule(lr){1-7}
\textbf{Overall} & 835 & 29.35 & 19.15 & 43.92 & 46.64 & 48.56 \\

\multicolumn{7}{c}{\textbf{MMLongBench-Doc}} \\
\textbf{Modality Group} & $n$ & G3-27B & G3-4B & GPT & Q3-30B & Q3-8B \\
\midrule
Image + Layout & 47  & 14.54 & 8.51  & 14.18 & 14.47 & 8.51 \\
Image + Text   & 92  & 18.48 & 13.94 & 23.45 & 26.18 & 21.26 \\
Image + Table  & 16  & 25.00 & 3.61  & 18.75 & 12.50 & 3.47 \\
Layout + Text  & 16  & 25.00 & 6.25  & 31.25 & 43.75 & 31.25 \\
Layout + Table & 9   & 11.11 & 19.66 & 11.11 & 33.33 & 11.11 \\
Text + Table   & 36  & 36.11 & 22.22 & 50.23 & 33.33 & 29.34 \\
\cmidrule(lr){1-7}
\textbf{Overall} & 216 & 21.22 & 13.04 & 25.61 & 25.41 & 18.83 \\
\bottomrule
\end{tabular}
\end{table}

\subsection{Analysis by LLM Model}\label{sec:model-analysis}

 \subsubsection{Descriptive Overview} 
 To compare modality contribution ($S$) and cooperation ($C$) across models, we macro-average results over the two benchmarks. Contribution scores average $S$ across all modality pairs containing a modality, while cooperation scores are reported at the modality-pair level (Figure~\ref{fig:aggregated_results_per_model_C_and_S}).
The contribution scores show that Table and Text provide the strongest marginal contributions, with Table contributing the most overall. In contrast, Layout contributes little as it is represented and retrieved in this pipeline, suggesting limited usefulness for QA. Image contributions vary more across models, with smaller models showing weaker Image contributions, indicating that extracting useful evidence from images may depend more strongly on model capacity.
Cooperation effects vary more strongly across models than contribution scores. Image-Text and Text-Table consistently exhibit negative cooperation, suggesting redundancy or modality dominance during answer generation. In contrast, modality pairs involving Table, particularly Image-Table and Layout-Table, show the strongest positive cooperation, although the magnitude varies across models.

\subsubsection{Bootstrap Analysis}

While the macro-averaged results described above offer a descriptive overview of the contribution and cooperation patterns across the different models, we complement them with a bootstrap-based analysis, details shown in Section~\ref{sec:metrics}, to quantify uncertainty and assess variability across different samples of questions. 

\begin{figure*}
  \centering
  \includegraphics[width=\linewidth]{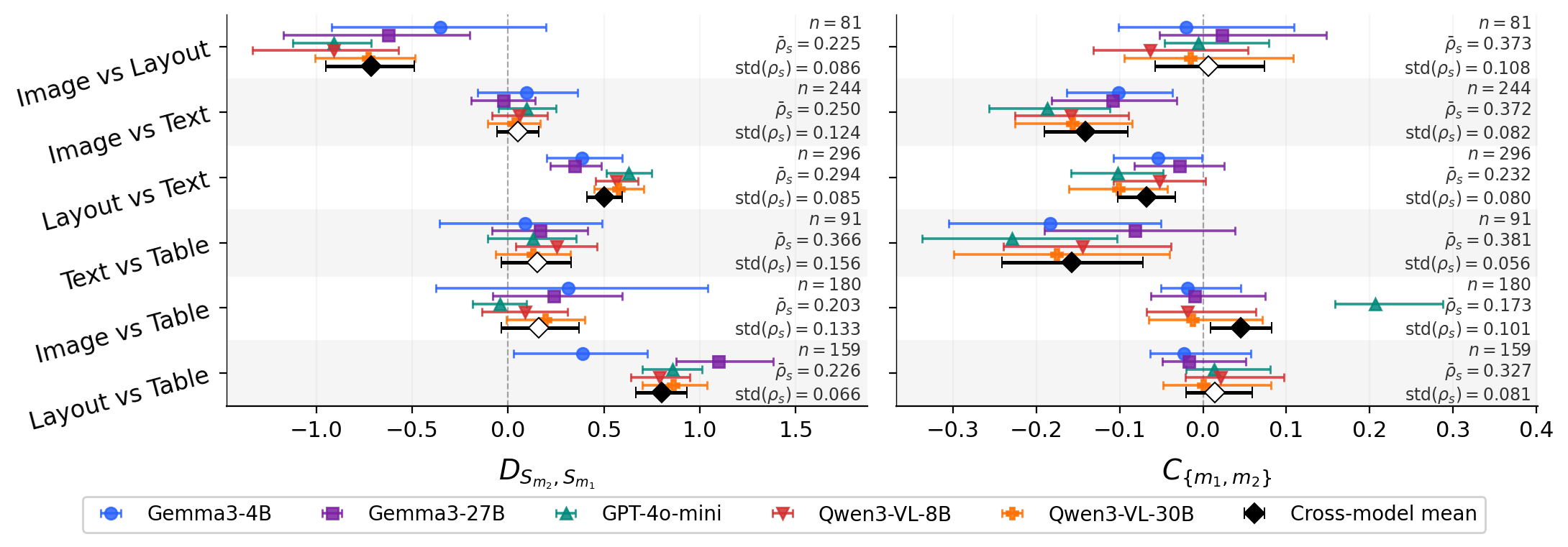}
  \caption{Overall cross-model contribution analysis (left) and cooperation analysis (right). Modality-pair scores are pooled across LongDocURL and MMLongBench-Doc using a benchmark-stratified bootstrap with 100,000 samples. Colored markers and horizontal intervals show model-specific bootstrap means with 95\% confidence intervals, while the black diamond denotes the cross-model mean, with a mean pairwise Spearman correlation $\bar{\rho}_s$ with standard deviation std($\bar{\rho}_s$). The value $n$ denotes the total number of questions available for sampling in each modality pair across both benchmarks.}
  \label{fig:cross_model_consistency_cooperation_contribution}
\end{figure*}

\paragraph{Modality contributions results}
Figure~\ref{fig:cross_model_consistency_cooperation_contribution} shows the bootstrap results for the contribution differences $D$. We can observe that Layout modality has a consistently lower contribution relative to the other modalities (see $D_{S_l,S_i}$, $D_{S_t,S_l}$, and $D_{S_{ta},S_l}$ in Figure~\ref{fig:cross_model_consistency_cooperation_contribution}), and this trend is broadly consistent across models, as reflected in positive average Spearman correlations. 
Table versus Text shows a larger contribution gap in favour of the Table modality, with stronger cross-model agreement as reflected by the comparatively higher $\bar{\rho}_s$ value. Text, however, contributes more than the Image modality in the Image+Text pair, but the gap is smaller, similarly to the gap between the Image and Table modality, where Tables consistently contribute more than Images.

\paragraph{Modality cooperation results.}
Figure~\ref{fig:cross_model_consistency_cooperation_contribution} shows the cooperation effects between modality pairs. Several pairs involving Text exhibit consistent redundancy. In particular, Image-Text and Text-Table show negative effects across models, with moderate cross-model agreement ($\bar{\rho}_s = 0.372$). A similar but weaker trend appears for Layout-Text. This suggests that textual information often overlaps with content available in other modalities, reducing the benefit of combining them. In contrast, Image-Layout and Layout-Table show effects close to zero, with confidence intervals overlapping zero, indicating little to no interaction between modalities. Evidence of complementary interactions is limited. The only pair showing slight cooperation is Image-Table, mainly driven by GPT-4o-mini, suggesting that this model may integrate image and table information more effectively.

\subsection{Analysis by Question Type}
\label{sec:questiontype}

In this section, we examine how modality contributions and cooperation vary across different question types, considering both question intent and question task requirements.

\subsubsection{Question Intent} 
Using the query-intent labels described in Section~\ref{sec:taxonomy}, we pool questions from both benchmarks according to their intent and recompute the SHAPE contribution and cooperation scores within each intent group and modality pair. Similarly to Section~\ref{sec:model-analysis}, we perform bootstrap resampling to estimate the uncertainty of these scores and to assess whether observed patterns are stable across different question samples (see Section~\ref{sec:metrics} for specifics regarding our bootstrap sampling scheme). The number of questions for each intent group is reported in Table~\ref{tab:intent_modality_counts}. No questions were assigned to the Language and Weather categories.

\begin{table}[width=.9\columnwidth]
\caption{Number of questions per intent and modality group. Modality groups are abbreviated as: T = Text, Ta = Table, I = Image, L = Layout.}
\label{tab:intent_modality_counts}
\begin{tabular}{lrrrrrr}
\toprule
\textbf{Intent} & L+T & I+T & T+Ta & I+Ta & L+Ta & I+L \\
\midrule
Entity       & 159 & 21  & 12 & 118 & 7  & 14 \\
List         & 41  & 43  & 14 & 39  & 131 & 17 \\
Quantity     & 39  & 94 & 37 & 13  & 11 & 37 \\
Verification & 26  & 36  & 9  & 2   & 3  & 4 \\
Attribute    & 13  & 19  & 5  & 1   & 2  & 4 \\
Calculation  & 4   & 11   & 12  & 5   & 3  & 4 \\
Description  & 6   & 6   & 1  & 0   & 0  & 1 \\
Location     & 1   & 3   & 0  & 1   & 2  & 0 \\
Opinion      & 1   & 2   & 0  & 1   & 0  & 0 \\
Temporal     & 3   & 4   & 0  & 0   & 0  & 0 \\
Process      & 1   & 3   & 0  & 0   & 0  & 0 \\
Reason       & 1   & 2   & 0  & 0   & 0  & 0 \\
Advice       & 1   & 0   & 0  & 0   & 0  & 0 \\
Resource     & 0   & 0   & 1  & 0   & 0  & 0 \\
\bottomrule
\end{tabular}
\end{table}

For this analysis, we group questions by both their intent and modality pair. This allows us to understand how the contribution and cooperation patterns change across different types of intent. Since the SHAPE scores calculated by groups with very small number of questions might be influenced by a few individual questions, we only include the groups that contain 10 or more questions.

\begin{figure*}
\centering

\includegraphics[width=0.59\textwidth]{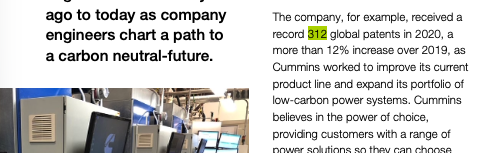}
\hspace{0.01\textwidth}
\includegraphics[width=0.37\textwidth]{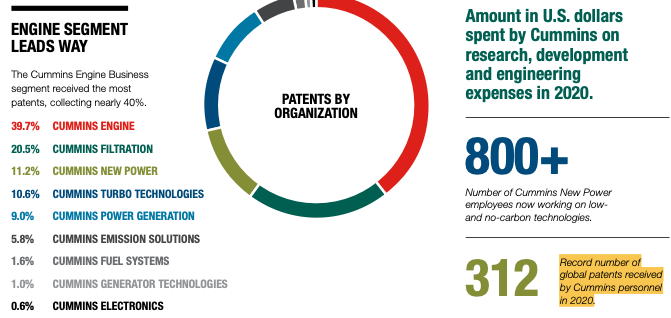}

\caption{Example where the relevant information appears in both text (left) and in an image (right). For the question “How many global patents did Cummins receive in 2020?", one out of five models answer accurately by using the Text modality, while three models answered correctly by using the Image modality.}
\label{fig:example}
\end{figure*}

\begin{figure*}
  \centering
  \includegraphics[width=\linewidth]{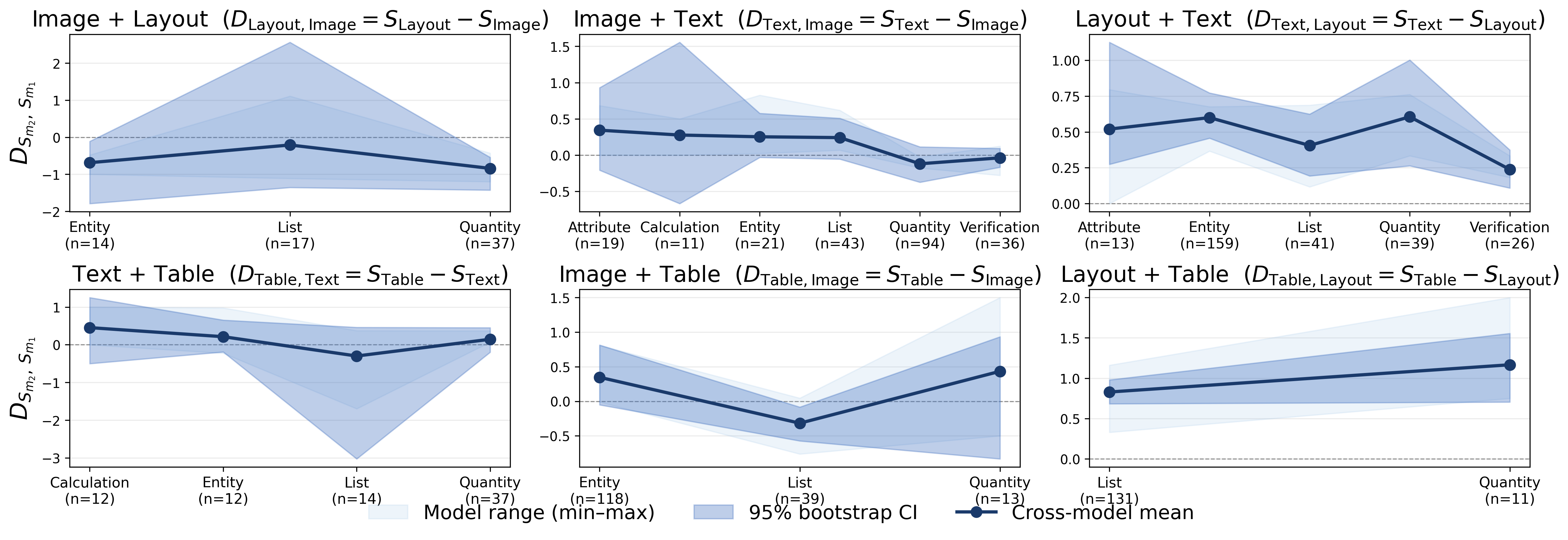}
  \caption{Contribution difference by question intent and modality pair. Each panel shows the mean contribution difference $D_{S_{m_2},S_{m_1}}$ across models for a specific modality pair. The darker shaded band indicates the 95\% bootstrap CI for the cross-model mean, and the lighter shaded band indicates variation across models.}
  \label{fig:aggregated_intent_D}
\end{figure*}

\paragraph{Contributions per intent.}
Figure~\ref{fig:aggregated_intent_D} shows that modality contributions vary across intents. Text consistently contributes more than Layout across all intents large enough to analyse, suggesting that layout information alone rarely provides primary evidence. For Quantity questions, the Text-Image and Table-Text pairs show contributions close to zero, indicating that neither modality consistently dominates. This may reflect cases where the answer is available in both modalities (Figure~\ref{fig:example}).

\begin{figure}
\centering
\includegraphics[width=0.41\textwidth]{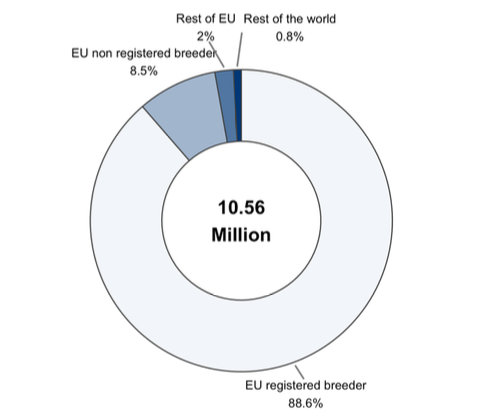}
\hfill
\includegraphics[width=0.47\textwidth]{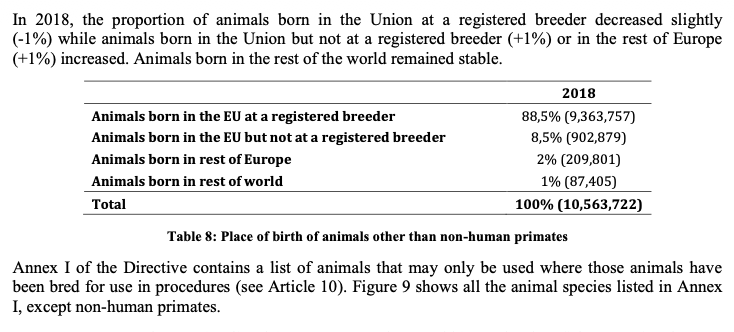}
\caption{Example where the relevant information appears in both the document image (top) and the table (bottom). For the question ``Give the distribution percentage of places of birth for animals used for the first time in 2018 (other than non-human primates)'', three out of five models answer accurately by using the image modality, while no model was able to answer correctly by using the table.}
\label{fig:list_intent}
\end{figure}

For List questions, Text contributes more than Image in the Text-Image pair, whereas Image contributes more in the Table-Image pair. We observe examples where models answer correctly using only visual information while failing with only table information (Figure~\ref{fig:list_intent}). Contributions in the Layout-Image and Table-Text pairs vary substantially across models, making stable conclusions difficult.

Entity questions exhibit more consistent patterns across models and bootstrap samples. Text and Table generally provide the strongest contributions, while the Table-Text pair shows little difference between modalities, likely reflecting redundancy because the relevant evidence is often available in both.

\begin{figure*}
  \centering
  \includegraphics[width=\linewidth]{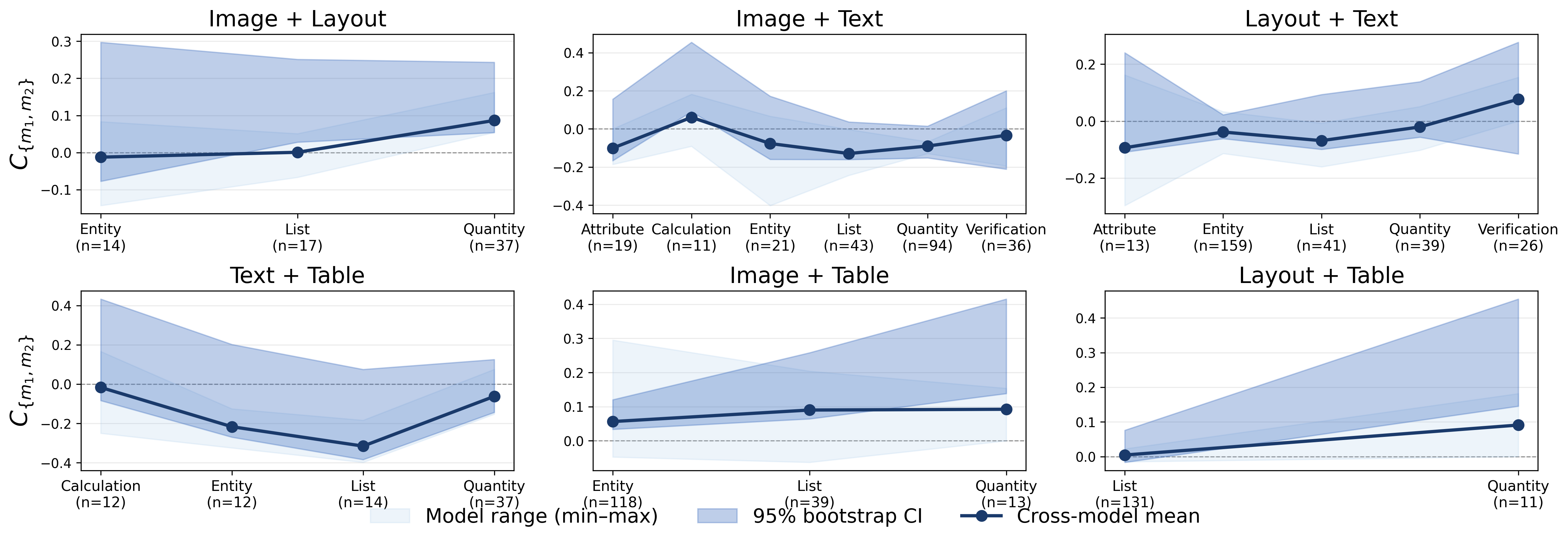}
  \caption{Modality cooperation by question intent and modality pair. Each panel shows the mean cooperation score $C_{\{m_1,m_2\}}$ across models for a specific modality pair. The darker shaded band indicates the 95\% bootstrap CI for the cross-model mean, and the lighter shaded band indicates variation across models.}
  \label{fig:aggregated_intent_C}
\end{figure*}

\paragraph{Cooperation per intent.}
Figure~\ref{fig:aggregated_intent_C} shows that cooperation varies across intents and modality pairs. For Quantity questions, Image-Layout, Image-Table, and Layout-Table pairs show positive cooperation, while pairs involving Text show negative or near-zero cooperation. Combined with the contribution analysis in Figure~\ref{fig:aggregated_intent_D} and Figure~\ref{fig:example}, this suggests that Quantity questions often contain redundant evidence across Text and other modalities. In contrast, modalities such as Image, Table, and Layout may each capture complementary aspects of the evidence, such as object counts, table structure, or spatial organization, leading to stronger cooperation. 
 
For List questions, modality pairs involving Text tend to show negative cooperation, suggesting limited benefit from combining modalities. In the Image-Text pair, Text often already provides sufficient information, while images mainly act as a filtering condition. A similar pattern appears for Entity questions, where pairs involving Text also show negative or negligible cooperation despite Text having the highest contribution. This suggests that Text alone is often sufficient to identify the correct entity without additional visual grounding.

\subsubsection{Question Task Requirement} 
In this section, we analyse how modality effects vary across different QA processes using the LongDocURL labels: Understanding, Reasoning, and Locating. Understanding questions require direct information extraction, Reasoning questions involve operations such as counting or comparison, and Locating questions require identifying relevant document elements through relational cues. This analysis uses the LongDocURL task-type labels and so it is restricted to the questions of that benchmark (n=835). As we did in Section~\ref{sec:questiontype}, we only analyse groups containing 10 or more questions. The number of questions for each question type group is reported in Table~\ref{tab:task_type_counts}.

\begin{table}[width=.9\columnwidth]
\caption{Number of questions per task type and modality group.}
\label{tab:task_type_counts}
\begin{tabular}{lrrr}
\toprule
\textbf{Modality group} & \textbf{Locating} & \textbf{Reasoning} & \textbf{Understanding} \\
\midrule
Image+Layout    & 0   & 11 & 23  \\
Image+Text    & 0   & 36 & 116 \\
Image+Table   & 144 & 6  & 14  \\
Layout+Text    & 137 & 22 & 121 \\
Layout+Table   & 126 & 4  & 20  \\
Text+Table   & 0   & 10 & 45  \\
\midrule
\textbf{Total} & \textbf{407} & \textbf{89} & \textbf{339} \\
\bottomrule
\end{tabular}
\end{table}

\begin{figure}
  \centering
  \includegraphics[width=\linewidth]{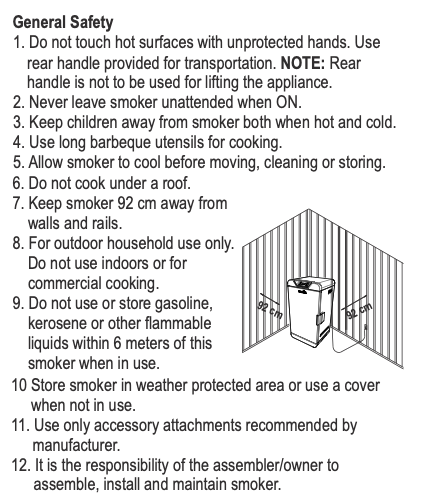}
  \caption{Example part of document where information appears in both textual and visual content. Given the question ``What should be the minimum distance (in cm) to the grill from the wall?'', all models fail to answer correctly based solely on the visual information.}
  \label{fig:understanding_task_example}
\end{figure}

\begin{figure*}
  \centering
  \includegraphics[width=\linewidth]{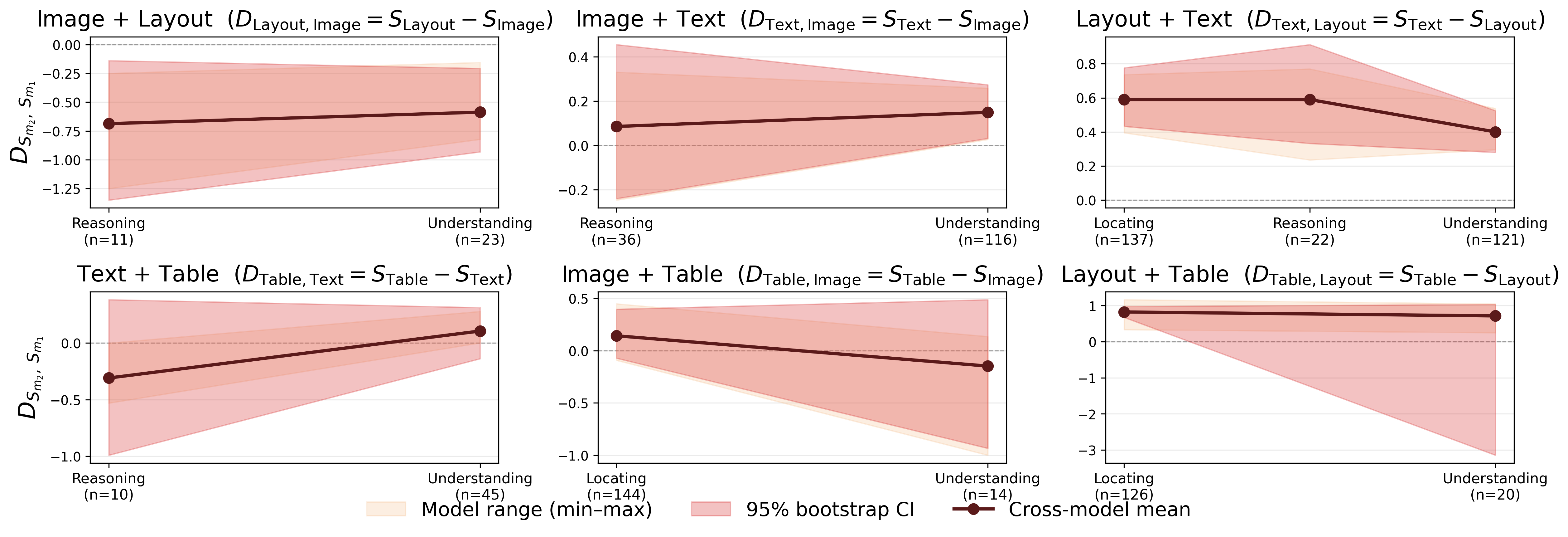}
  \caption{Contribution difference by task type and modality pair. Each panel shows the cross-model mean contribution difference $D_{S_{m_2},S_{m_1}}$ for a given modality pair, grouped by task type. The darker shaded band indicates the 95\% bootstrap CI for the cross-model mean, and the lighter shaded band indicates variation across models.}
  \label{fig:aggregated_tasktype_D}
\end{figure*}

\paragraph{Contributions per question task requirement.}
Figure~\ref{fig:aggregated_tasktype_D} shows that modality contributions vary by task type. For Understanding questions, Text contributes more in Text-Image and Text-Layout pairs, while Table dominates in Table-Text pairs, suggesting that text supports direct extraction whereas tables are more useful for dense structured information. Figure~\ref{fig:understanding_task_example} shows an example case where all models failed to extract the relevant information from the image. For Locating questions, Text and Table contribute most, reducing the usefulness of Layout. For Reasoning questions, Text generally contributes more in Table-Text and Text-Layout pairs, suggesting that models reason more effectively over textual descriptions than tables.

\begin{figure*}
  \centering
  \includegraphics[width=\linewidth]{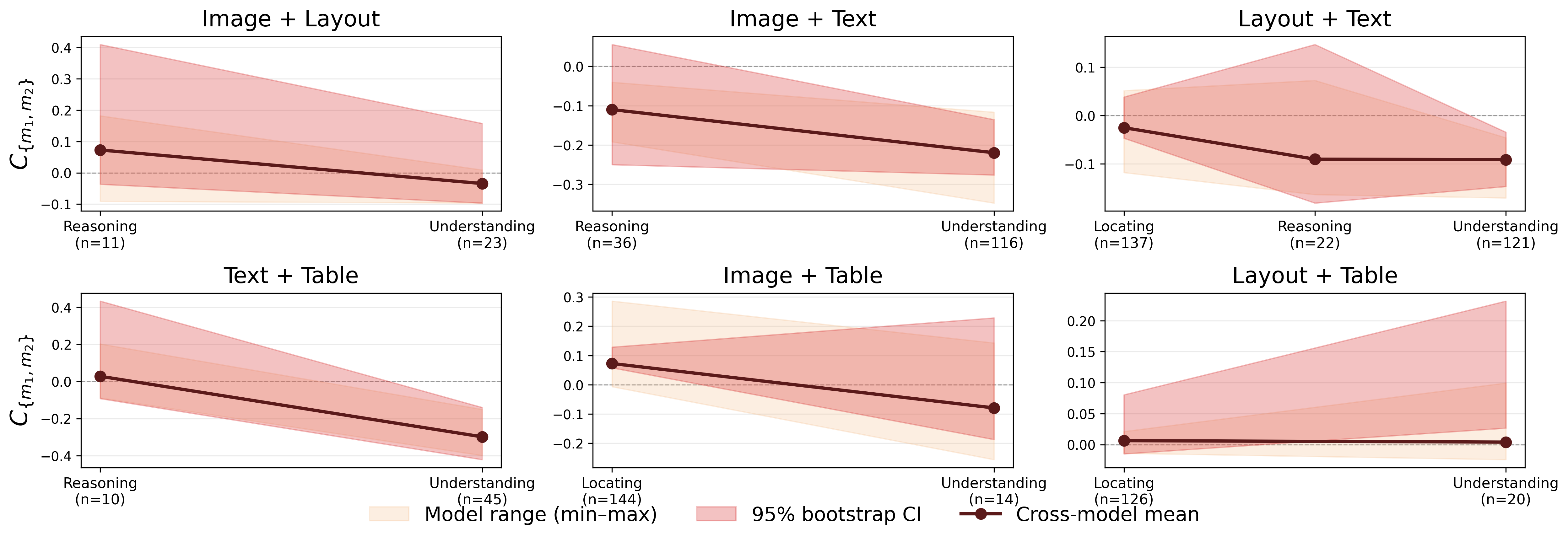}
  \caption{Modality cooperation by task type and modality pair. Each panel shows the cross-model mean cooperation score $C_{\{m_1,m_2\}}$ for a given modality pair, grouped by task type. The darker shaded band indicates the 95\% bootstrap CI for the cross-model mean, and the lighter shaded band indicated variation across models.}
  \label{fig:aggregated_tasktype_C}
\end{figure*}

\paragraph{Cooperation per question task requirement.}
Figure~\ref{fig:aggregated_tasktype_C} shows that Understanding questions exhibit lower cooperation for the Image-Layout, Image-Text, and Text-Table pairs compared to Reasoning questions. This suggests that direct extraction tasks often require only one modality, making additional modalities redundant or even harmful. In contrast, Locating questions show positive cooperation for the Image-Table pair. Although Table usually contributes more, the two modalities appear complementary: the table provides the relevant information, while the image helps identify its location on the page. 

\section{Discussion}

The findings in our work are important for multimodal retrieval in GraphRAG systems, as well as for deciding which modalities to integrate and use for a given application. In this section we first discuss about the findings of our analysis and common patterns identified through the experiments. Then, based on these observations, we provide design recommendations for multimodal GraphRAG systems.

\subsection{Observations}

\begin{figure}
  \centering
  \includegraphics[width=\linewidth]{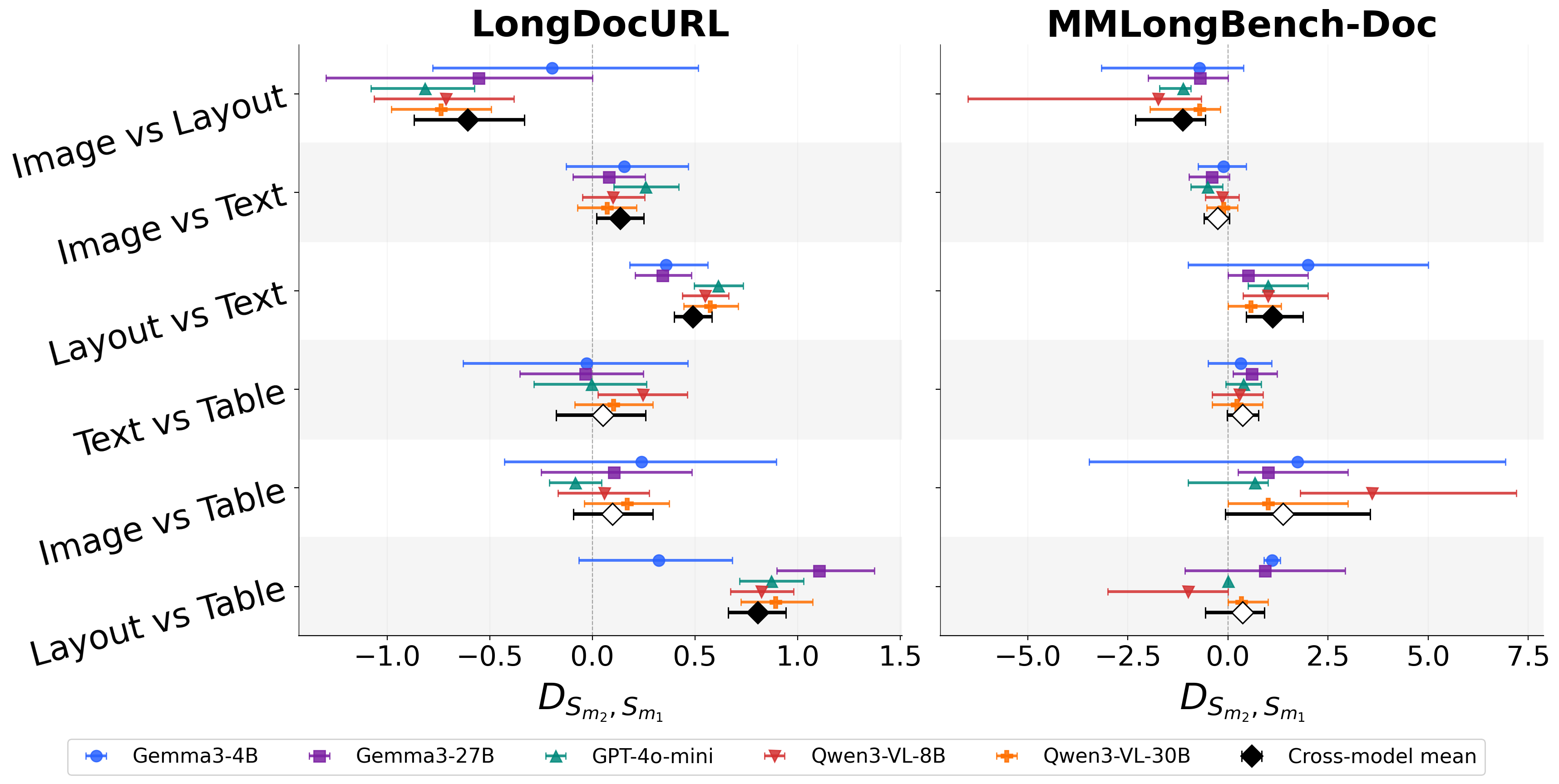}
  \caption{Modality-pair SHAPE contribution differences ($D$) estimated separately for LongDocURL (left) and MMLongBench-Doc (right) using 100,000 bootstrap samples.}
  \label{fig:contrib_comparison}
\end{figure}

\textbf{Contributions are not equal, with Tables and Text showing the strongest effects.} Across both benchmarks and all tested models, Text and Table modalities provide the strongest marginal contribution in terms of question answering performance (Figure~\ref{fig:cross_model_consistency_cooperation_contribution}). The Layout modality consistently shows the least contribution effect across every model, modality pair, question intent group and task type, in the large majority of model–pair combinations. Layout is also negative is several modality pairs in both benchmarks. Contribution effect of the Image modality varies and depends heavily on the evaluated model with smaller models generally extracting less visual information from images. While on LongDocURL Text shows a higher contribution compared to Image, on MMLongBench-Doc the effect is the oppostive for all models. This can be seen in Figure~\ref{fig:contrib_comparison}.

\textbf{Cooperation effects are weak and mostly non-text.} When averaging the findings over all questions (see Figure~\ref{fig:cross_model_consistency_cooperation_contribution}), cooperation is negative for pairs involving Text, and limited in non-text pairs. Only Image-Layout and Layout-Table show effects close to zero, and only the Image-Table pair shows positive cooperation mainly driven by a single model (GPT-4o-mini). At the level of cross-model means, this asymmetry is systematic as every negative effect we see in the figure involves the Text modality, while pairs without Text show neutral or positive cooperation. This non-synergistic effect can be attributed to the fact that the modalities supply overlapping rather than complementary evidence (Figure~\ref{fig:example}). This finding is consistent with prior work that shows that adding redundant context, even when factually correct, can distract the language model and decreases accuracy on questions the model could otherwise answer correctly~\cite{hsia2025ragged,luo2025zero}. For instance, authors in ~\cite{luo2025zero} propose pruning redundant knowledge already mastered by the model from the retrieval corpus, to improve retrieval efficiency and downstream performance. Our results further suggest that this redundancy varies systematically across modality pairs, question intents, and task types.

\textbf{Modality cooperation interactions depend on question intent and task type.} While the macro-averaged results showed mostly negative cooperation effects, grouping the questions by intent and task type shows that these effects are not uniform across all intents and task types. Looking at Figure~\ref{fig:aggregated_intent_C} and the Quantity questions, Image-Layout, Image-Table and Layout-Table all show positive cooperation effects, whereas in the pooled analysis (Figure~\ref{fig:cross_model_consistency_cooperation_contribution}) the same pairs show effects close to zero, with only Image+Table showing a slight positive effect driven mainly by GPT-4o-mini. For the same question intent group, we see that for the modality groups that contain Text the cooperation remains at zero or below it. Therefore, cooperation for modality pairs that involve Text stays at or below zero in both the pooled analysis and this intent group. 
A similar pattern can be seen when looking at specific task types. For example, for Locating questions the Image-Table pair shows a positive cooperation, while for the same modality pair, looking at Figure~\ref{fig:cross_model_consistency_cooperation_contribution} we can see a mostly negative or neutral effect once the GPT-4o-mini model is excluded. These results show that cooperation can be conditional and it appears for specific question intents and task types, and it is not based on a modality pair.  

\textbf{The value of visual evidence depends on the inference model.} Our findings indicate that across question types, the Image modality does not usually provide the dominant evidence. While Text and Table show the highest contribution scores, the contribution of images varies and as seen in Figure~\ref{fig:aggregated_results_per_model_C_and_S}, the contribution of the Image is lower for smaller models. This suggests that models with fewer parameters cannot utilise the visual information provided by the image as effectively as the models with more parameters. This indicates that the value of visual evidence in multimodal GraphRAG is a reflection of how capable a model is rather than a guarantee provided by the modality itself.

\subsection{Design Recommendations}

The observations above show that the usefulness of a modality, in terms of its contribution and cooperation effects, is not an intrinsic property. Rather, it depends on the other modalities that are retrieved alongside it, the question being asked, and the model used for inference. These observations have direct implications for how multimodal GraphRAG systems should determine which modalities to retrieve for DocVQA tasks. In the following paragraphs we provide a set of design recommendations based on our findings.

\textbf{Contribution and cooperation for adaptive modality selection in multimodal GraphRAG.} Retrieval systems must decide which modality or combination of modalities to retrieve for a given question. Existing approaches typically learn this decision by estimating which retrieval strategy yields the highest task performance~\cite{pmlr-v322-zhao26a}. Although this identifies the best-performing strategy, it does not explain why it performs best. Good performance may come from one modality containing information that other modalities lack, or from multiple modalities containing largely overlapping information. Each case has different implications for retrieval. In the first case, retrieving additional modalities can provide complementary evidence when the primary modality is insufficient for question answering. In the second case, retrieving multiple modalities can provide largely redundant information and increase retrieval cost without adding evidence. Our analysis distinguishes between these two cases by measuring both the contribution of individual modalities and their cooperation, revealing when multiple modalities provide complementary information and when they are largely redundant. We further identify factors that give rise to this redundancy, providing insights into when multimodal retrieval is likely to be beneficial. 
These findings have implications for adaptive retrieval systems such as multimodal GraphRAG. Rather than selecting modalities solely on the basis of expected task performance, multimodal GraphRAG systems could also exploit contribution and cooperation to determine whether an additional modality is likely to provide complementary evidence or merely redundant information, thereby improving downstream efficiency and performance. 

\textbf{Multimodal GraphRAG should link entities and relations to their source modalities, enabling modality-aware retrieval.} Our results show that the contribution and cooperation effects of a modality are not fixed and largely depend on the question intent, the task type, and the inference model. Systems that do not selectively retrieve evidence from specific modalities cannot accommodate this variation. Multimodal GraphRAG is a natural fit for this, because evidence from different modalities is transformed into entities and relations~\cite{guo2025rag,bu2025query}. By recording the modality of each piece of evidence on the graph, every retrieved fact can be associated with the modality that supports it. We therefore argue that multimodal GraphRAG systems should link entities and relations to their source modalities during the graph construction phase and offer the option for selective retrieval based on a set of modalities. This enables retrieval policies that suppress or prioritise evidence based on its modality. Our extension of RAG-Anything demonstrates that this linking can be integrated into the existing pipeline without requiring modifications to other components, while introducing only modest overhead.

\textbf{Retrieval recommendations on DocVQA.} Our analysis shows that the usefulness of modality pairs depends on the intent of a question, and the type of reasoning required to answer it. However, some patterns are stable across benchmarks, models, question intent, and task groups. First, prioritising the retrieval of Text and Table is a safe practical default as they usually offer the strongest marginal contributions across almost all settings. Second, the Layout modality is the weakest contributor and thus it should be deprioritised as a source of evidence. Third, Text should not be expected to exhibit strong synergistic effects when combined with another modality. Across the pooled analysis, as well as nearly all task types and question intents, Text consistently shows negative or near zero cooperation. Consequently, retrieving an additional modality is only worthwhile when Text alone does not provide sufficient information. Lastly, positive cooperation is primarily observed between non-text modalities. Image-Table exhibits the most consistent positive cooperation, particularly for Quantity and Locating questions. These findings suggest that adaptive retrieval systems should selectively retrieve non-text modalities when they are expected to provide complementary evidence, rather than treating multimodal retrieval as the default.

\section{Scope and Generalisability}

\textbf{Choice of GraphRAG pipeline.} We conducted our analysis using the multimodal GraphRAG framework, RAG-Anything~\cite{guo2025rag}, a state-of-the-art framework that explicitly models both textual and non-textual content through a unified graph representation while preserving modality-specific information during graph construction. Different GraphRAG frameworks differ in graph construction, indexing, and retrieval, and prior work~\cite{ju-etal-2025-mire,fine-grained-late} has shown that multimodal retrieval performance is sensitive to these variations. Thus, we mitigate sensitivity to framework-specific retrieval behaviour (see description below). Moreover, our analyses focus on relative modality contributions and cooperation effects rather than absolute retrieval performance. Investigating the extent to which the observed patterns generalise across frameworks is beyond the scope of this work.

\textbf{Sensitivity to retrieval design.} To reduce the dependence of the results on the retrieval mechanism, we adapt the retrieval strategy of RAG-Anything to first filter on modality-specific chunks before ranking them by embedding cosine similarity.  The same ranking procedure is applied across all modality configurations, ensuring the observed differences arise from the modalities available for retrieval rather than from changes in the retrieval process itself. We further retrieve the top-20 most relevant chunks, favouring recall over precision so that relevant evidence is captured even when ranking is imperfect. For LongDocURL we also use the largest context window supported by each model to minimise truncation of retrieved evidence. Together, these choices aim to reduce the influence of framework-specific retrieval behaviour, allowing our analyses to focus on modality contribution and cooperation effects.

\textbf{LLM-based intent annotation.} The question intents were assigned by using an LLM-based approach, and validated by two LLM judges rather than by human annotators. At least one judge suggested a different label for 170 out of 1,051 questions (16.2\%). Out of those 170 questions, 66 were revised to a label that both judges agree on, while for 97 of those questions (9.2\%) only one judge disagreed with the original label. Only 7 questions (0.7\%) received a original label that both judges disagreed on without agreeing on a replacement label. The mislabeled questions are most likely to affect the smaller question intent groups in which questions that are wrongly assigned a label can shift the SHAPE scores. The larger question intent groups (Entity, List, Quantity) are more robust to a small number of mislabeled questions. 

\section{Conclusion}
In this paper we presented a modality-aware KG-based question answering framework for analysing contribution and interaction of text, table, image and layout modalities in DocVQA. Through experiments across five multimodal LLMs, two long-context DocVQA benchmarks, and using a state-of-the-art multimodal GraphRAG pipeline, we showed that tables and text often provide the strongest evidence, while additional modalities do not consistently improve performance and can introduce redundancy. Our findings further demonstrate that modality interactions depend strongly on the question intent and task type, with complementary effects appearing when modalities provide distinct forms of evidence. Our findings argue for selective, modality-aware retrieval in the design of more effective GraphRAG systems, where modalities are filtered according to the downstream task rather than retrieved uniformly.

\printcredits

\section*{Declaration of generative AI and AI-assisted technologies in the manuscript preparation process.}

During the preparation of this work the authors used Claude in order to paraphrase and reword. After using this tool/service, the authors reviewed and edited the content as needed and take full responsibility for the content of the published article.

\section*{Declaration of competing interest.}

The authors declare that they have no known competing financial interests or personal relationships that could have appeared to influence the work reported in this paper.

\section*{Data availability}

The implementation code for the framework and the analysis is publicly available at: https://github.com/Antonis-Georgakopoulos/mmkg-modality-analysis

\bibliographystyle{cas-model2-names}

\bibliography{cas-refs}





\end{document}